\documentclass[twocolumn,tighten,times]{aastex62}

\usepackage[figuresright]{rotating}

\newcommand{\grp}[1]{\textbf{\textit{#1}}}

\graphicspath{{./}{figures/}}

\accepted{for publication in ApJ}

\shorttitle{Neighboring Galaxies' Hydrodynamic Impact on Star Formation}
\shortauthors{Moon, An, \& Yoon}

\begin{document}

\title{Living with Neighbors. I. Observational Clues to Hydrodynamic Impact of Neighboring Galaxies on Star Formation}

\correspondingauthor{Suk-Jin Yoon}
\email{sjyoon0691@yonsei.ac.kr}

\author[0000-0001-7075-4156]{Jun-Sung Moon}
\affil{Department of Astronomy, Yonsei University, Seoul, 03722, Republic of Korea}
\affil{Center for Galaxy Evolution Research, Yonsei University, Seoul, 03722, Republic of Korea}

\author[0000-0003-3791-0860]{Sung-Ho An}
\affil{Department of Astronomy, Yonsei University, Seoul, 03722, Republic of Korea}
\affil{Center for Galaxy Evolution Research, Yonsei University, Seoul, 03722, Republic of Korea}

\author[0000-0002-1842-4325]{Suk-Jin Yoon}
\affil{Department of Astronomy, Yonsei University, Seoul, 03722, Republic of Korea}
\affil{Center for Galaxy Evolution Research, Yonsei University, Seoul, 03722, Republic of Korea}



\begin{abstract}

Galaxies in pairs show enhanced star formation (SF) compared to their counterparts in isolation, which is often explained by the tidal effect of neighboring galaxies. Recent observations, however, reported that galaxies paired with early-type neighbors do not undergo the SF enhancement. Here we revisit the influence of neighbors using a large sample of paired galaxies from the Sloan Digital Sky Survey and a carefully constructed control sample of isolated counterparts. We find that star-forming neighbors enhance SF, and even more so for more star-forming (and closer) neighbors, which can be attributed to collisions of interstellar medium (ISM) leading to SF. We further find that, contrary to the anticipated tidal effect, quiescent neighbors quench SF, and even more so for more quiescent (and closer) neighbors. This seems to be due to removal of gas reservoirs via ram pressure stripping and gas accretion cutoff by hot gas halos of quiescent neighbors, on top of their paucity of ISM to collide to form stars. Our findings, especially the intimate connection of SF to the status and strength of neighbors' SF, imply that the hydrodynamic mechanisms, along with the tidal effect, play a crucial role during the early phase of galactic interactions.

\end{abstract}




\section{Introduction} \label{sec:intro}

Interactions between galaxies are a common phenomenon. About 1\%$-$5\% of galaxies in the local universe are observed to be interacting with other companions of similar mass \citep[major interactions; e.g.,][]{2008ApJ...681..232L, 2010MNRAS.401.1043D, 2011ApJ...742..103L, 2012ApJ...747...85X, 2015A&A...576A..53L}. The observed fraction of paired galaxies rapidly increases with redshift and reaches about 10\,\% at $z$ $>$ 1 \citep[e.g.,][]{2016ApJ...830...89M, 2017A&A...608A...9V}. When taking into consideration that minor interactions with smaller satellites are $\sim$\,3 times more frequent than major interactions \citep[e.g.,][]{2011ApJ...742..103L, 2014MNRAS.440.2944K, 2015MNRAS.452.2845K}, a galaxy is expected to interact with another galaxy every few gigayears. Hence, the evolutionary history of galaxies has an inseparable link with galaxy--galaxy interactions.

One of the remarkable features found in interacting galaxies is the enhancement of star formation (SF) activity. Many observations have reported that the SF rate (SFR) in close galaxy pairs is higher than isolated galaxies \citep{2000ApJ...530..660B, 2003MNRAS.346.1189L, 2012A&A...539A..45L, 2004MNRAS.352.1081A, 2004MNRAS.355..874N, 2006AJ....132..197W, 2010AJ....139.1857W, 2007AJ....134..527W, 2008AJ....135.1877E, 2013MNRAS.435.3627E, 2008MNRAS.385.1903L, 2010MNRAS.401.1552D, 2011MNRAS.412..591P, 2013MNRAS.433L..59P, 2011ApJ...728..119W,  2014MNRAS.437.2137S}. The theoretical expectation agrees well with the observational results \citep{2006MNRAS.373.1013C, 2008MNRAS.384..386C, 2007A&A...468...61D, 2008A&A...492...31D, 2013MNRAS.433L..59P, 2015MNRAS.448.1107M}. For example, \citet{2013MNRAS.433L..59P} compared the interaction-induced SF in galaxy pairs of the Sloan Digital Sky Survey \citep[SDSS;][]{2000AJ....120.1579Y} with hydrodynamic merger simulations and found a general agreement with the predicted SFR enhancement as a function of the separation between paired galaxies out to $\sim$\,200 kpc.

While the SFR enhancement of paired galaxies is primarily explained by the tidal effect between galaxies, some recent studies have focused on the role of the hydrodynamic effect. \citet{2009ApJ...691.1828P} investigated, using the SDSS, the impact of interactions on galaxy properties such as the color and SFR depending on the morphology of interacting neighbors. They found that only late-type neighbors induce SF, while early-type neighbors suppress the SF activity. After this, several observational studies of paired galaxies in infrared wavelengths revealed similar trends. \citet{2010A&A...522A..33H} suggested that the fraction of luminous infrared galaxies and ultraluminous infrared galaxies increases only near to late-type neighbors. \citet{2011A&A...535A..60H} also showed that the SFR based on the infrared luminosity increases as the distance to the nearest neighbor decreases, particularly when the neighbor is a late-type galaxy. \citet{2010ApJ...713..330X} observed 27 close major-merger pairs using $Spitzer$ observations and also concluded that only spiral galaxies with a spiral neighbor show the SFR enhancement compared to the control sample, while spirals with an elliptical neighbor do not. All the studies above took notice of the impact of hot gas halos around the early-type neighboring galaxies, such as ram pressure stripping and cutoff of the cold gas supply, speculating that the removal of the gas reservoir can suppress further SF.

Recently, \citet{2016ApJS..222...16C} studied 88 close major-merger pairs observed with $Herschel$ and confirmed again that only spiral+spiral (S+S) pairs show the SFR enhancement compared to the isolated control sample, unlike spiral+elliptical (S+E) pairs. They also found that the ratio of total dust mass to stellar mass is not clearly distinct between the S+S and S+E pairs, claiming that the SF efficiency (i.e., the SFR per unit gas mass) of the S+S pairs is higher than that of the S+E pairs. \citet{2016ApJ...829...78D} showed an enhancement in the intensity of the interstellar radiation field for S+S pairs, while S+E pairs did not show a clear difference from the control sample. \citet{2018ApJS..237....2Z} obtained the H \textsc{\romannumeral 1} observation of 70 pairs from the sample of \citet{2016ApJS..222...16C} and found that the S+S pairs, compared to the S+E pairs, have a similar H \textsc{\romannumeral 1} gas fraction but a higher SF efficiency of cold gas. They argued that the same gas fraction between the S+S and S+E pairs does not support the scenario in which the gas supply is cut off by the shock-heated hot halos and instead proposed that the collision between the interstellar media (ISMs) in two spiral galaxies enhances the SFR. This result may imply that a certain kind of hydrodynamic effect plays a role in inducing SF within interacting galaxy pairs.

It is, however, not trivial to assess the exact effect of interactions in the presence of the selection bias. For instance, since galaxies in pairs are preferentially selected in denser environments, they are to be less star-forming than isolated galaxies \citep[e.g.,][]{2004MNRAS.353..713K}, and the selection effect may hide the SF enhancement by interactions. Another difficulty lies in the dependence of the SF enhancement on multiple parameters, such as the mass ratio between two galaxies \citep{2006AJ....132..197W, 2008MNRAS.384..386C, 2008AJ....135.1877E}, the local environment \citep{2010MNRAS.407.1514E}, and the orbital orientation \citep{2007A&A...468...61D, 2008A&A...492...31D, 2013MNRAS.433L..59P}. For this reason, a solid conclusion can be drawn only from the statistical approach. Fortunately, the advent of large surveys such as the SDSS has enabled us to make use of a large sample of paired galaxies.

In this series of papers, we explore both observationally and theoretically the characteristics of galactic interactions with neighbors and their impact on the properties and evolution of galaxies in terms of stellar populations and dynamics. In the present Paper I, we examine observationally the effect of the nearest neighbor on the SF activity based on a large sample of paired galaxies from the SDSS and a careful control on the selection bias. We in particular focus on the hydrodynamic interplay between the target galaxies and their interacting neighbors. As an indicator of the hydrodynamic gas properties of galaxies in pairs, we use their SFRs{\footnote {\citet{2015PKAS...30..469M} previously used the morphology of galaxies to conduct a similar analysis, but the morphology is susceptible to interactions and often difficult to determine for close pairs that are of primary interest in this paper. Another choice would be a more direct indicator such as the cold gas fraction in galaxies. But the number of galaxies with measured cold gas contents is limited. With an intent to maximize the number of objects in the sample, we assume the SFR represents the hydrodynamic state of galaxies.}} that are estimated from the optical emission lines. That is to say, the main concern of this paper is to examine the role of the neighbors' SF properties in the interaction-induced SF. Most previous studies on this topic have not considered the SF activities of neighbors. Even though some pioneering work has been carried out, these studies did not use either a strict control sample \citep[e.g.,][]{2009ApJ...691.1828P} or a large enough sample to draw a general conclusion \citep[e.g.,][]{2016ApJS..222...16C}. We here intend to fill this gap in this paper.

The paper is organized as follows. In Section \ref{sec:sample}, we describe the data collection and the sample selection. Then, in Section \ref{sec:control}, we construct the control sample to ensure a valid comparison between the paired and isolated galaxy samples. We present the main results in Section \ref{sec:result}, which is followed by a discussion in Section \ref{sec:discuss}. Finally, we conclude in Section \ref{sec:conclusion}. We adopt the standard $\Lambda$CDM cosmology with $\Omega_m$ = 0.3, $\Omega_\Lambda$ = 0.7, $H_0$ = 100 $h$\,km\,s$^{-1}$Mpc$^{-1}$, and $h$ = 0.7 throughout the paper.

\section{Sample Selection} \label{sec:sample}

\subsection{Data} \label{subsec:data}

In order to compile a large data sample of interacting galaxies, we use the SDSS Data Release 7 \citep{2009ApJS..182..543A}, which provides spectroscopic data of nearly 1 million galaxies magnitude-limited to $r$ $\leq$ 17.77 \citep{2002AJ....124.1810S}. Our interacting galaxy sample is selected among galaxies with a high-confidence spectroscopic redshift (\texttt{SpecObjAll.zConf} $>$ 0.7) to reduce the contamination by spurious pairs. The sample galaxies also should be classified as `Galaxy' based on both photometry (\texttt{PhotoObjAll.type} = 3) and spectroscopy (\texttt{SpecObjAll.specClass} = 2). These criteria for compilation are similar to those of \citet{2013MNRAS.433L..59P}, who also utilized a massive sample of galaxy pairs to investigate their SF activities. We restrict our data to galaxies lying in the redshift range of 0.02 $<$ $z$ $<$ 0.1, where a reliable measurement of the local density is possible with enough completeness \citep{2006MNRAS.373..469B}. The stellar masses of galaxies are estimated by \citet{2014ApJS..210....3M} using the broadband SED fitting with dusty models and the improved photometry of \citet{2011ApJS..196...11S}. The catalog contains 326,833 galaxies with reliable redshift, stellar mass, and local environment measurements. 

The SFRs of galaxies are brought in from the MPA-JHU catalog \citep{2004MNRAS.351.1151B}, where the SFRs are derived from the optical emission lines such as H$\alpha$, H$\beta$, [O \textsc{\romannumeral 2}], [O \textsc{\romannumeral 3}], [N \textsc{\romannumeral 2}], and [S \textsc{\romannumeral 2}] with proper consideration of the dust attenuation. We use the specific SFR (sSFR), i.e., SFR per unit stellar mass, measured within the SDSS 3$''$ fiber, instead of the aperture-corrected global SFR. The enhancement of SF by galaxy interactions is known to be concentrated at the central region of galaxies, as confirmed by both the observations \citep[e.g.,][]{2011MNRAS.412..591P, 2015A&A...579A..45B} and numerical simulations \citep[e.g.,][]{2008MNRAS.384..386C, 2015MNRAS.448.1107M}. We also take it into consideration that applying an empirical aperture correction for merging galaxies can produce an undesired bias. It is worth noting that the physical diameter of the fiber at 0.02 $<$ $z$ $<$ 0.1 varies from 0.9 to 3.9 $h^{-1}$kpc, yet we have proper controls over the redshift and stellar mass of galaxies (see Section \ref{sec:control}), which guarantees a fair comparison of the fiber SFR between the pair and control samples.

All AGN candidates are not used in our analysis to isolate the change in the SFR from the contamination from the AGN activity. According to previous studies \citep{2007MNRAS.375.1017A, 2007AJ....134..527W, 2011MNRAS.418.2043E, 2012A&A...538A..15H}, the interaction with neighboring galaxies affects the AGN activity, and the radiation from AGNs can be confused with H$\alpha$ fluxes coming from star-forming regions. The AGN candidates are identified by using the empirical lines of \citet{2003MNRAS.346.1055K} on the BPT diagram \citep{1981PASP...93....5B}. We exclude 67,906 AGN candidates ($\sim$\,20\,\% of galaxies) from our sample that are classified as ``composite", ``AGN," or ``low S/N AGN" \citep{2004MNRAS.351.1151B}. 

\subsection{Neighbor Identification} \label{subsec:neighbor}

To identify paired galaxies, we search a nearest galaxy candidate for each galaxy within a loose interval of radial velocities ($\pm$1000 km\,s$^{-1}$). In this step, we only take into account galaxies with the stellar mass larger than 1\,/\,10 of that of the target. If a galaxy does not have any neighbor within the projected radius of 200 $h^{-1}$kpc, we call it an ``isolated" galaxy and set it aside for a later use. 

Among the remaining galaxies, we select our pair sample with the following criteria: (a) The nearest neighbor should be within the projected physical separation of 200 $h^{-1}$kpc from the target galaxy. (b) We also require the minimum separation of 3$''$ to reduce the blending of galactic light. (c) The relative radial velocity between the target and the nearest neighbor should be less than 300 km s$^{-1}$. The relative radial velocity cut of 300 km\,s$^{-1}$ is a compromise between the size and quality of the pair sample, which has been used in many previous studies \citep[e.g.,][]{2013MNRAS.433L..59P}. (d) The stellar mass ratio between the target and the nearest neighbor should be between 0.1 and 10, and thus two galaxies involved in the interaction have roughly comparable masses. (e) We also require that the nearest neighbor should exert the largest tidal influence on the target. The tidal influence $\Theta_{\mathrm{nei}}$ is defined, for each neighbor, as
\begin{equation}
\Theta_{\mathrm{nei}} = \frac{M_{\mathrm{*,nei}}}{r_{\mathrm{nei}}^3},
\end{equation}
where $M_{\mathrm{*,nei}}$ and $r_{\mathrm{nei}}$ are the stellar mass and physical projected distance to the neighbor, respectively. The purpose of this criterion is to minimize the influence of another perturber. (f) We restrict our analysis to galaxies with a separation to the survey boundary greater than 4 $h^{-1}$Mpc.

Isolated and paired galaxies identified in the SDSS spectroscopic survey may have additional neighbors that are fainter than its limiting magnitude of $r$ = 17.77. Besides, due to fiber collision, the SDSS spectroscopic sample suffers from incompleteness for galaxy pairs closer than 55$''$ on the sky \citep{2008ApJ...685..235P}. In order to obtain a purer sample, we search through the SDSS photometric redshift catalog \citep{2003AJ....125..580C, 2007ApJS..172..634A} and exclude galaxies that violate our selection criteria. In particular, we take out $\sim$\,100,000 otherwise-isolated galaxies that have a photometric neighbor closer than 200 $h^{-1}$kpc within the photometric redshift uncertainty. We also exclude $\sim$\,20,000 paired galaxies that have yet another photometric neighbor with a closer projected distance or a larger tidal influence within the photometric redshift uncertainty. The stellar mass for the photometric objects is derived by the formula of \citet{2003ApJS..149..289B}. We note that the uniqueness of our sample selection lies in that each galaxy in the sample must identify a single neighbor. The above criterion ($e$) guarantees that the neighbor is the most influential one to the target galaxy. While the sample selection is basically designed to be consistent with previous studies  \citep[e.g.,][]{2013MNRAS.433L..59P}, adopting the criteria, along with the fact that galaxies do not have additional photometric neighbors, is suited for our analysis. We also check the images of close paired galaxies and further exclude 89 galaxies because they are in fact single galaxies identified as two objects owing to the presence of dust lanes or bright spots on their disks. As a consequence, we are left with 33,182 isolated galaxies and 14,432 paired galaxies that meet the all criteria described above. 

\subsection{Classification of Paired Galaxies} \label{subsec:classification}

Galaxies in the pair sample are classified into two groups according to the SF activity of their interacting neighbor. Many studies on the interaction-induced SF have not considered the SF activity of the neighbor \citep[e.g.,][]{2013MNRAS.433L..59P}. As a demarcation between quiescent and star-forming galaxies, we use log(sSFR) $=$ $-$11.5 (yr$^{-1}$), where the valley of the sSFR distribution is located. We refer to the groups of target galaxies interacting with a quiescent neighbor and a star-forming neighbor as the ``\grp{q}" group and the ``\grp{s}" group, respectively. We identify 7071 galaxies in the \grp{q} group and 7361 galaxies in the \grp{s} group.

\begin{figure*}[t]
\epsscale{0.842}
\plotone{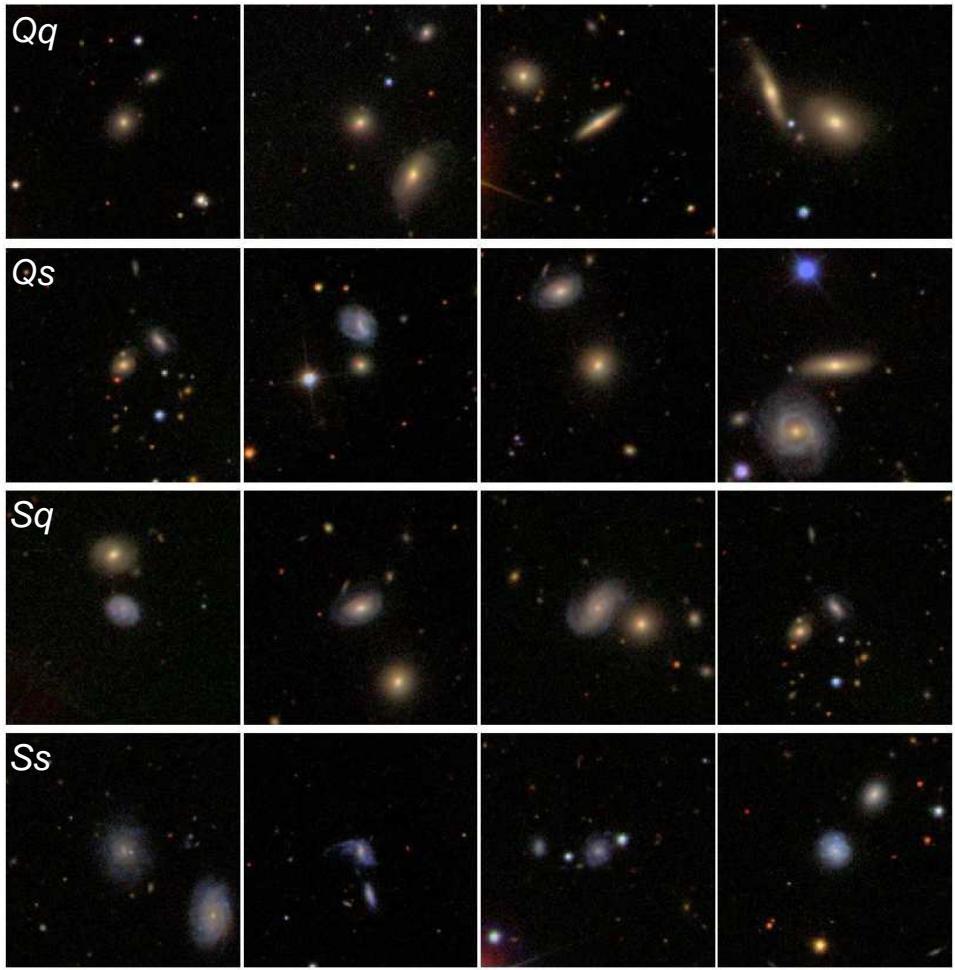}
\caption{Examples of galaxies in our pair sample. The paired galaxies are at the center of stamp images. The pair sample is further classified into four groups, \grp{Qq} (first row), \grp{Qs} (second row), \grp{Sq} (third row), and \grp{Ss} (fourth row), according to the SF activities of the target and those of the nearest neighbor. In each group, four examples are shown among paired galaxies having a neighbor within a projected separation of 30 $h^{-1}$kpc. Each image is 2$'$ on a side. \label{fig:imgpair}}
\end{figure*}

\begin{figure*}[t]
\epsscale{0.87}
\plotone{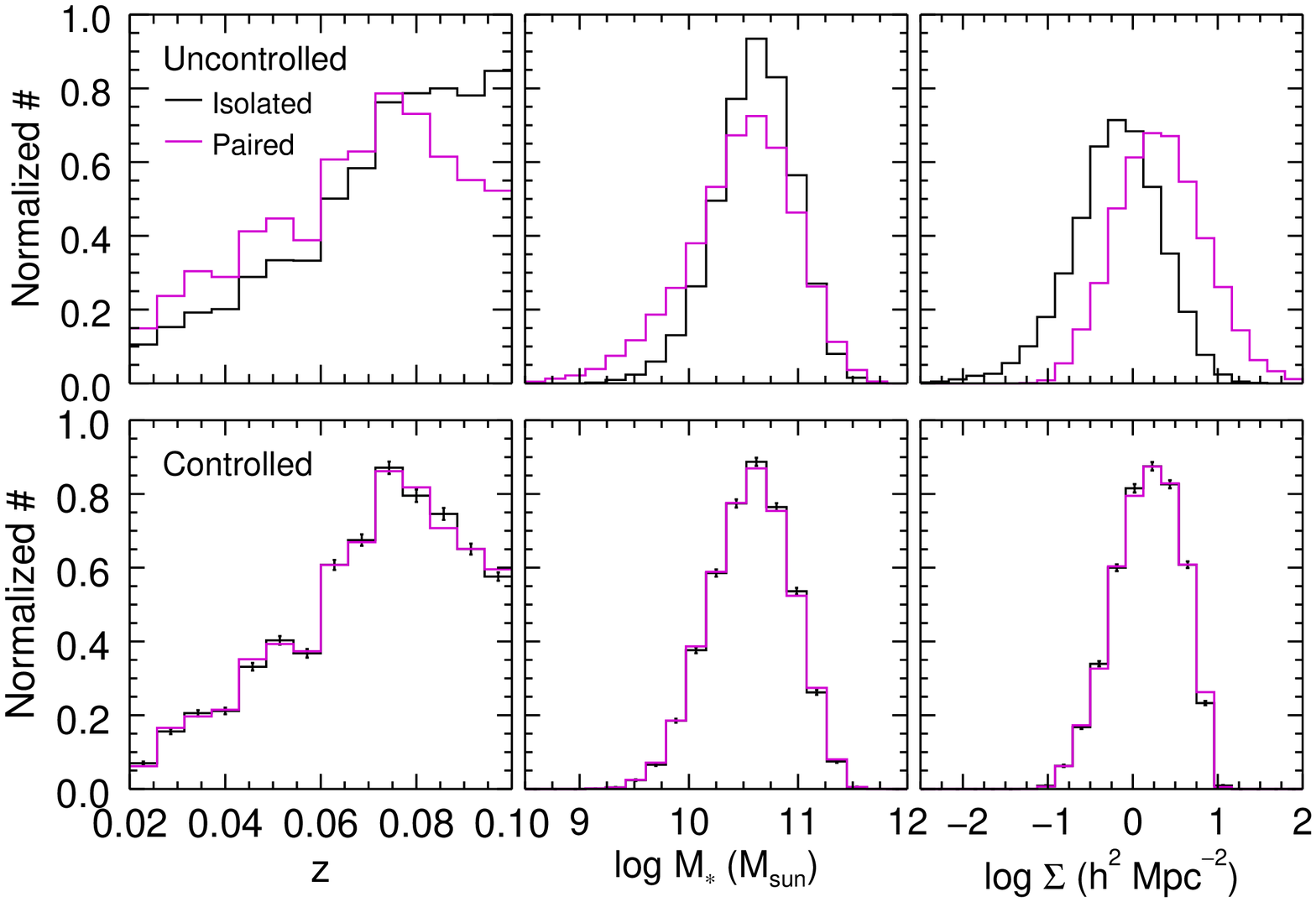}
\caption{Top: distributions of the redshift (left), stellar mass (middle), and local density (right) for all isolated galaxies (black) and all paired galaxies (purple). Bottom: The same as the top row, but for the controlled galaxies (see Section~\ref{sec:control}). Error bars show the standard deviation obtained from 1000 random selections in the control sample. Each distribution is normalized to the number of galaxies it contains. \label{fig:histcont}}
\end{figure*}

Additionally, the target galaxy is expressed as uppercase letters: ``\grp{Q}" as a quiescent target and ``\grp{S}" as a star-forming target. The uppercase letters do not necessarily mean that target galaxies are more massive than their neighbors denoted by the lowercase letters. The ``\grp{Qq}" group consists of quiescent target galaxies interacting with a quiescent neighbor, and the ``\grp{Qs}" group consists of quiescent target galaxies interacting with a star-forming neighbor. Likewise, the ``\grp{Sq}" and ``\grp{Ss}" groups are star-forming galaxies interacting with a quiescent neighbor and a star-forming neighbor, respectively. We identify 3848, 2459, 3223, and 4902 galaxies in the \grp{Qq}, \grp{Qs}, \grp{Sq}, and \grp{Ss} groups, respectively. Figure \ref{fig:imgpair} displays the sample galaxies of each group.

\section{Control Sample} \label{sec:control}

To examine the influence of interacting neighbors, one needs to construct a proper control sample first. The procedure to select paired galaxies inevitably introduces some undesired biases. For instance, galaxies in pairs tend to reside preferentially in denser environments, where galaxies are on average more passive and redder \citep{2007ApJ...671.1538B, 2009MNRAS.397..748P, 2009MNRAS.399.1157P, 2011MNRAS.412..591P}. This selection bias can be misinterpreted as an interaction-induced effect when the comparison is done between the isolated and paired galaxy samples without careful consideration of a control sample.

We construct the control sample using the isolated galaxies selected in Section \ref{subsec:neighbor}. The isolated galaxies do not have any neighbor closer than the projected distance of 200 $h^{-1}$kpc within a radial velocity range of $\pm$1000 km\,s$^{-1}$. By randomly selecting an isolated galaxy for each paired galaxy within the range of 0.005 in redshift, 0.1 dex in stellar mass, and 0.1 dex in local density, we make the distributions of the redshift, stellar mass, and local density of the control sample identical to those of the paired galaxy sample. \citet{2009MNRAS.397..748P} tested biases for the pair sample using the mock galaxy catalog based on the Millennium simulation \citep{2005Natur.435..629S}, and suggested that an unbiased control sample can be made by forcing the distributions of the redshift, stellar mass, local density, bulge-to-total ratio, and halo mass to be identical to those of the pair sample. We follow a similar approach in this study, but there are some differences in the details. First, we do not use the dark matter halo mass because it is observationally very hard to estimate in practice. \citet{2009MNRAS.399.1157P} pointed out that the halo mass seems less significant in the real galaxy data than what they had earlier found in the simulation. Second, we do not use any morphological information such as the bulge-to-total ratio and the S\'{e}rsic index. Because tidal interactions modify the galaxy morphology, it is not guaranteed that these parameters describe the pre-interaction state of paired galaxies. Consequently, our approach to the control sample is similar to that of \citet{2013MNRAS.433L..59P, 2016MNRAS.461.2589P}, who used the redshift, stellar mass, and local density parameters.

The local density used for making the control sample is calculated by the adaptive kernel approach \citep{1986desd.book.....S}. The adaptive kernel estimator assigns the kernel bandwidth to each galaxy based on the initial density estimate with a fixed-width kernel. The method has the advantage of being less scale dependent than other popular density estimators. Besides, statistical fluctuations in the estimated density are smoothed by using a larger kernel size in a lower-density region, while subtle density structures in high-density regions are well captured. The adaptive kernel estimator is also known to outperform other estimators in tests with a simulated density field \citep{2011A&A...531A.114F, 2015ApJ...805..121D}. The procedure is as follows:
\begin{enumerate}

\item Calculate a pilot density estimate for the $i$th galaxy $\hat{\Sigma}(\textbf{r}_i)$ with a fixed kernel bandwidth. We use the projected number density of galaxies within a radial velocity range of $\pm$1000 km\,s$^{-1}$ from the target as the initial pilot estimate. For $N$ galaxies within the slice including the $i$th galaxy
\begin{equation}
\hat{\Sigma}(\textbf{r}_i) = \sum_{j \neq i}^N K(|\textbf{r}_i-\textbf{r} _j|, w),
\end{equation}
where $K(|\textbf{r}_i-\textbf{r} _j|, w)$ is the kernel function and $w$ is the fixed bandwidth. We choose the kernel bandwidth $w$ as 1 $h^{-1}$Mpc. The standard 2D Gaussian kernel is defined as
\begin{equation}
K(|\textbf{r}_i-\textbf{r} _j|, w) = \frac{1}{2 \pi w^2}\exp(-\frac{|\textbf{r}_i-\textbf{r} _j|^2}{2 w^2}).
\end{equation}

\item Define the local bandwidth factor $\lambda_i$ as a function of the pilot estimate
\begin{equation}
\lambda_i = (\hat{\Sigma}_i/g)^{-0.5},
\end{equation}
where $g$ is the geometric mean of the pilot estimates of all galaxies within a radial velocity range of $\pm$1000 km\,s$^{-1}$ from the $i$th galaxy. The local bandwidth factor $\lambda_i$ gets smaller as the galaxy is located in a denser region.

\item Compute the adaptive kernel estimate $\Sigma(\textbf{r}_i)$ at the location of galaxies with a new adaptive kernel bandwidth $\lambda_j w$:
\begin{equation}
\Sigma(\textbf{r}_i) = \sum_{j}^N K(|\textbf{r}_i-\textbf{r} _j|, \lambda_j w).
\end{equation}
When estimating the local density, we make no attempt to correct for the redshift dependence, since in this study we do not compare the local density parameter between galaxies at different redshifts. 

\end{enumerate}

\begin{figure*}[t]
\epsscale{1.17}
\plotone{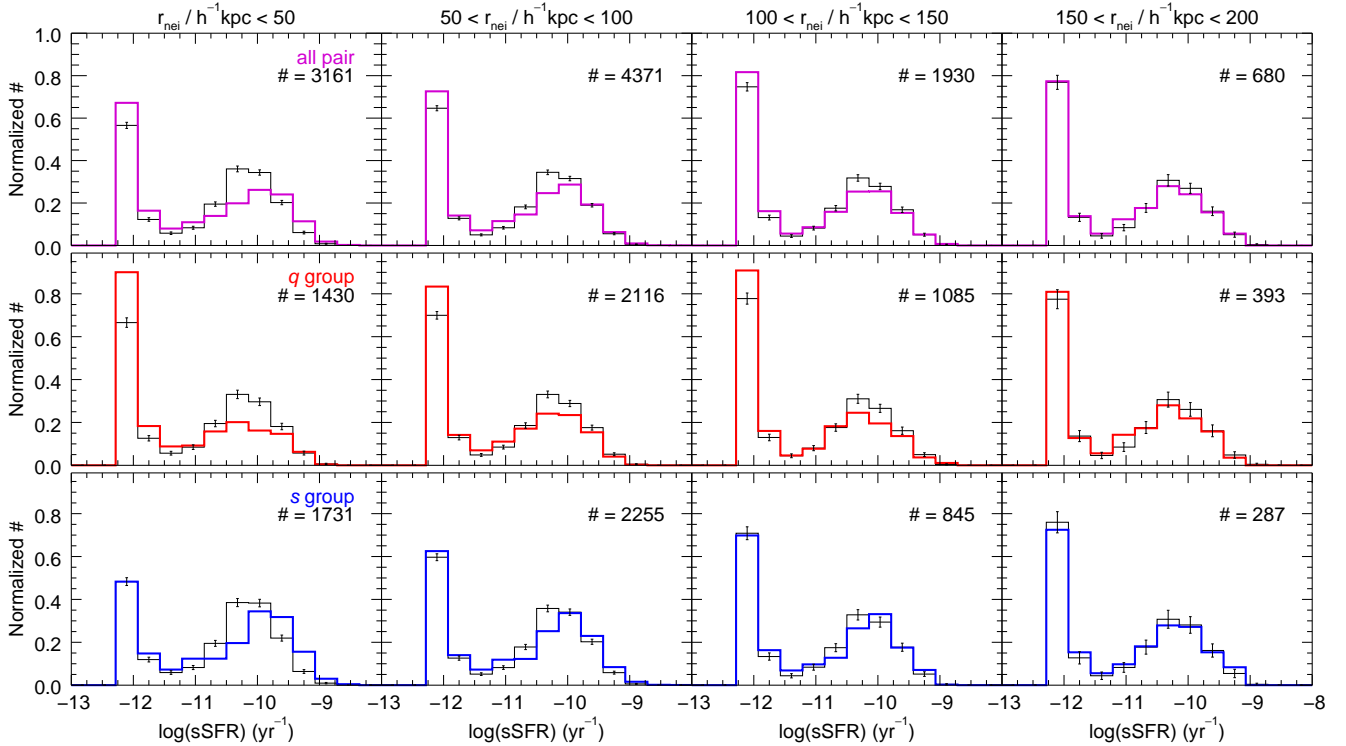}
\caption{sSFR distributions for the whole pair sample (purple), the \grp{q} group (red), and the \grp{s} group (blue). From left to right, the distance to the nearest neighbor increases. The distributions of the control samples are shown as black histograms. Error bars show the standard deviation obtained from 1000 random selections in the control samples. Each distribution is normalized to the number of galaxies it contains. \label{fig:ssfrhist}}
\vspace{0.1cm}
\end{figure*}

Figure \ref{fig:histcont} compares the redshift, stellar mass, and local density of the isolated and paired galaxies for the uncontrolled sample (top row) and the controlled sample (bottom row). The top panels show that, compared to the isolated galaxies, the paired galaxies tend to be at a lower redshift, of a lower mass, and in a denser environment. The differences in the distributions come from the selection effect, rather than the real nature of the isolated and paired galaxy samples. The redshift difference is likely due to the SDSS fiber collision. A lower-mass galaxy has a higher chance of having a roughly comparable mass neighbor thanks to the teeming population of low-mass galaxies. Besides, a galaxy in a denser region is more apt to have a neighbor.

The bottom panels of Figure \ref{fig:histcont} show the distributions for the controlled samples. We generate 1000 random selections in the isolated control sample and plot the standard deviation as error bars. In the ideal case, the isolated control sample should have the same distributions as the original pair sample, but in practice some paired galaxies do not have any isolated counterpart within the given range of 0.005 in redshift, 0.1 dex in stellar mass, and 0.1 dex in local density. Besides, for better statistics, we also restrict our pair sample to galaxies whose number of isolated counterparts is greater than 10. This condition eliminates paired galaxies especially located in dense environments where isolated counterparts are rare. The purpose of this is to prevent repetitive selection of the same isolated galaxy into our control sample, which can distort the statistics of isolated galaxies, particularly when the sample size is small. The median number of the isolated counterparts for the paired galaxies in our final sample is 68, and using different values for the minimum number of counterparts does not change our overall conclusion. The number of paired galaxies having corresponding control counterparts is 10,142, among which 5024 are in the \grp{q} group and 5118 are in the \grp{s} group.

\section{Results} \label{sec:result}

Figure \ref{fig:ssfrhist} shows the sSFR distribution of paired galaxies as a function of the projected separation to the nearest neighbor. From left to right, the distance to the nearest neighbor increases. In the first row, the sSFR distribution for the entire sample is more dispersed than the control sample; the portions of galaxies in both the quiescent (log(sSFR) $<$ $-$11.5) and actively star-forming (log(sSFR) $>$ $-$9.5) modes increase with respect to the control sample as the separation decreases. The influence of interactions seems to start at $r_{\mathrm{nei}}$ $\sim$ 150 $h^{-1}$kpc and becomes more evident at $r_{\mathrm{nei}}$ $<$ 50 $h^{-1}$kpc. This result is in line with previous work that showed that the optical color distribution of galaxies in pairs has an excess in both red and blue tails compared to the control galaxies \citep{2009MNRAS.399.1157P, 2010MNRAS.401.1552D}. The excess of the blue part has been interpreted as the interaction-induced enhancement of SF. For the excess of the red part, a tidally induced dusty SF or old stellar ages due to gas stripping were suggested as possible origins \citep{2009MNRAS.399.1157P}.

\begin{figure*}[t]
\epsscale{0.85}
\plotone{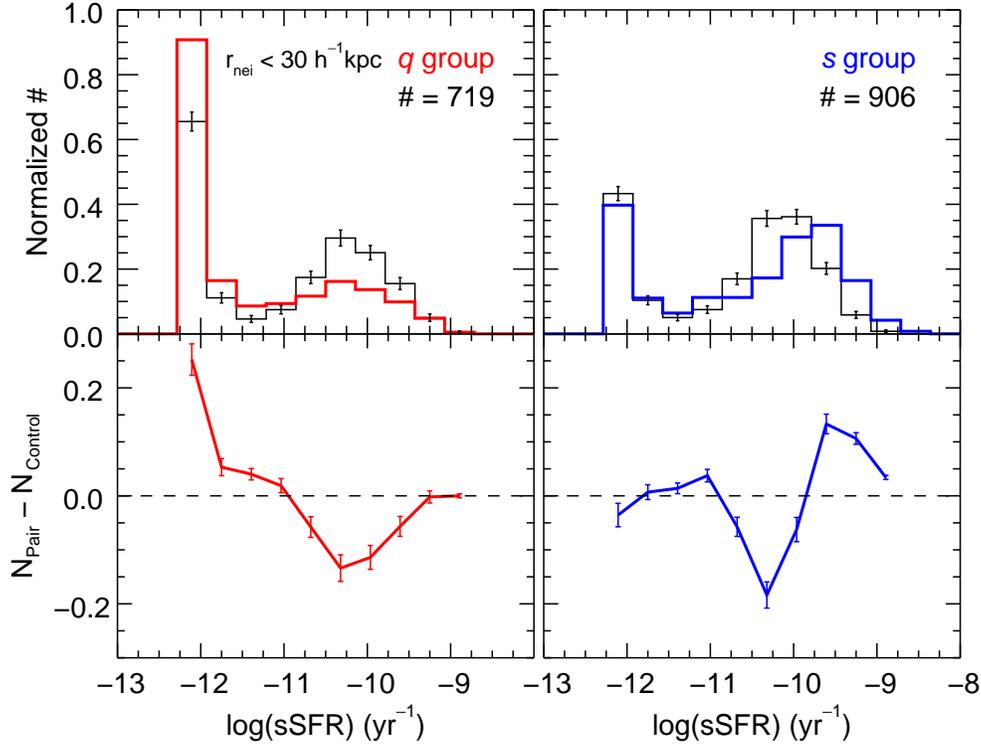}
\caption{Top: sSFR distributions for the \grp{q} group (left panel, red) and the \grp{s} group (right panel, blue) with a projected separation less than 30 $h^{-1}$kpc. The distributions of the control sample are shown as black histograms. Error bars show the standard deviation obtained from 1000 random selections in the control samples. Bottom: The difference in number between the pair and control samples ($N_\mathrm{pair}$$-$$N_\mathrm{control}$) for each sSFR bin. \label{fig:ssfrhist30}}
\end{figure*}

When dividing the paired galaxy sample into the \grp{q} (second row) and \grp{s} (third row) groups, we find that the two groups show remarkably different trends. The \grp{q} group has a higher fraction of quiescent galaxies than the control sample. The \grp{q} group no longer exhibits the sSFR enhancement that is observed in the whole pair sample. Contrary to the \grp{q} group, the \grp{s} group has a higher mean sSFR compared to the control sample, with no change in the number fraction of star-forming galaxies. Figure \ref{fig:ssfrhist30} highlights the behavior of the paired galaxies at a small separation to the neighbor. We choose $r_{\mathrm{nei}}$ $<$ 30 $h^{-1}$kpc for this figure. The sSFR distributions (top panels) and the difference in the number between the pair and control samples at each sSFR bin (bottom panels) are shown for the \grp{q} (left column) and \grp{s} (right column) groups. The major difference between the \grp{q} and \grp{s} groups is evident at both extremes; the increase of the quiescent galaxies (log(sSFR) $<$ $-$11.5) is led by the \grp{q} group, and the increase of actively star-forming galaxies (log(sSFR) $>$ $-$9.5) is led by the \grp{s} group.

In short, we find that the effect of interactions depends on the SF activity of the interacting neighbor. The effect on the sSFR shown in Figures \ref{fig:ssfrhist} and \ref{fig:ssfrhist30} is twofold: (1) for the \grp{q} group, the number fraction of star-forming galaxies decreases, and (2) for the \grp{s} group, there is an enhancement of the mean SFR for star-forming galaxies. We will scrutinize the two aspects more closely in the following subsections.

\subsection{Interaction-induced ``Reduction'' in the Star-forming Galaxy Number Fraction of the \grp{q} Group} \label{subsec:sffrac}

\begin{figure*}[t]
\epsscale{0.95}
\plotone{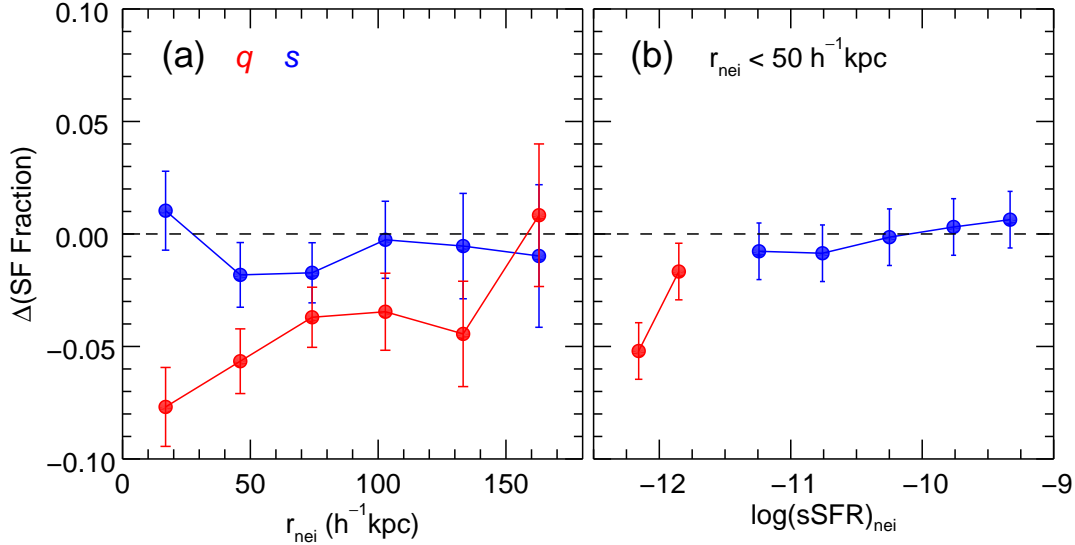}
\caption{Interaction-induced changes in the number fraction of star-forming galaxies for the \grp{q} (red) and \grp{s} (blue) groups as functions of (a) the projected distance to the nearest neighbor and (b) the sSFR of the nearest neighbor. A minus value means that the number of star-forming galaxies decreases more than expected from the control samples. For each bin, 1000 random selections in the control samples are generated, and the mean value is plotted along with the standard error. \label{fig:sffrac}}
\end{figure*}

Figure \ref{fig:sffrac} shows the effect of interactions on the number fraction of star-forming galaxies in the \grp{q} (red) and \grp{s} (blue) groups. In each bin for $r_{\mathrm{nei}}$ and log(sSFR)$_{\mathrm{nei}}$, the pair sample and the control sample contain an equal number of galaxies that have nearly identical distributions of redshift, stellar mass, and local density. Hence, the difference can be interpreted as the consequence of interactions. The interaction-induced change in the number fraction of star-forming galaxies is defined as
\begin{equation}
\Delta(\mathrm{SF~Fraction}) = \frac{N_{\mathrm{SF,\,pair}} - N_{\mathrm{SF,\,control}}}{N_{\mathrm{Tot}}},
\end{equation}
where $N_{\mathrm{Tot}}$ is the total number of galaxies in the whole pair sample, $N_{\mathrm{SF,\,pair}}$ is the number of star-forming galaxies in the pair samples (i.e., the \grp{q} [red] and \grp{s} [blue] groups), and $N_{\mathrm{SF,\,control}}$ is the number of star-forming galaxies in the control sample corresponding to the pair sample. We generate the random control sample 1000 times and calculate the mean values and standard errors of $\Delta$(SF Fraction). A minus value means that the SF of galaxies is quenched by the neighbor.

Figure \ref{fig:sffrac}(a) shows the trend as a function of $r_{\mathrm{nei}}$. The most remarkable feature is that the \grp{q} group has a low possibility of being star-forming. The decrease of the star-forming fraction in the \grp{q} group is extended to 150 $h^{-1}$kpc in separation. The star-forming fraction of the \grp{s} group remains similar to the control sample even when they have a very close star-forming companion. Figure \ref{fig:sffrac}(b) shows the trend as a function of sSFR of the nearest neighbor. We restrict the sample to galaxies with a separation of less than 50 $h^{-1}$kpc to see a clear trend. The reduction of star-forming galaxy fraction is only found in the \grp{q} group, and the reduction is severe for the \grp{q} galaxies having fully quenched neighbors (i.e., log(sSFR) $<$ $-$12.0). The \grp{s} group does not show a significant deviation from the control sample, which implies that the interactions cannot rejuvenate fully quenched galaxies, at least, during the early phase of the interaction prior to the coalescence.

\begin{figure*}[t]
\epsscale{1.0}
\plotone{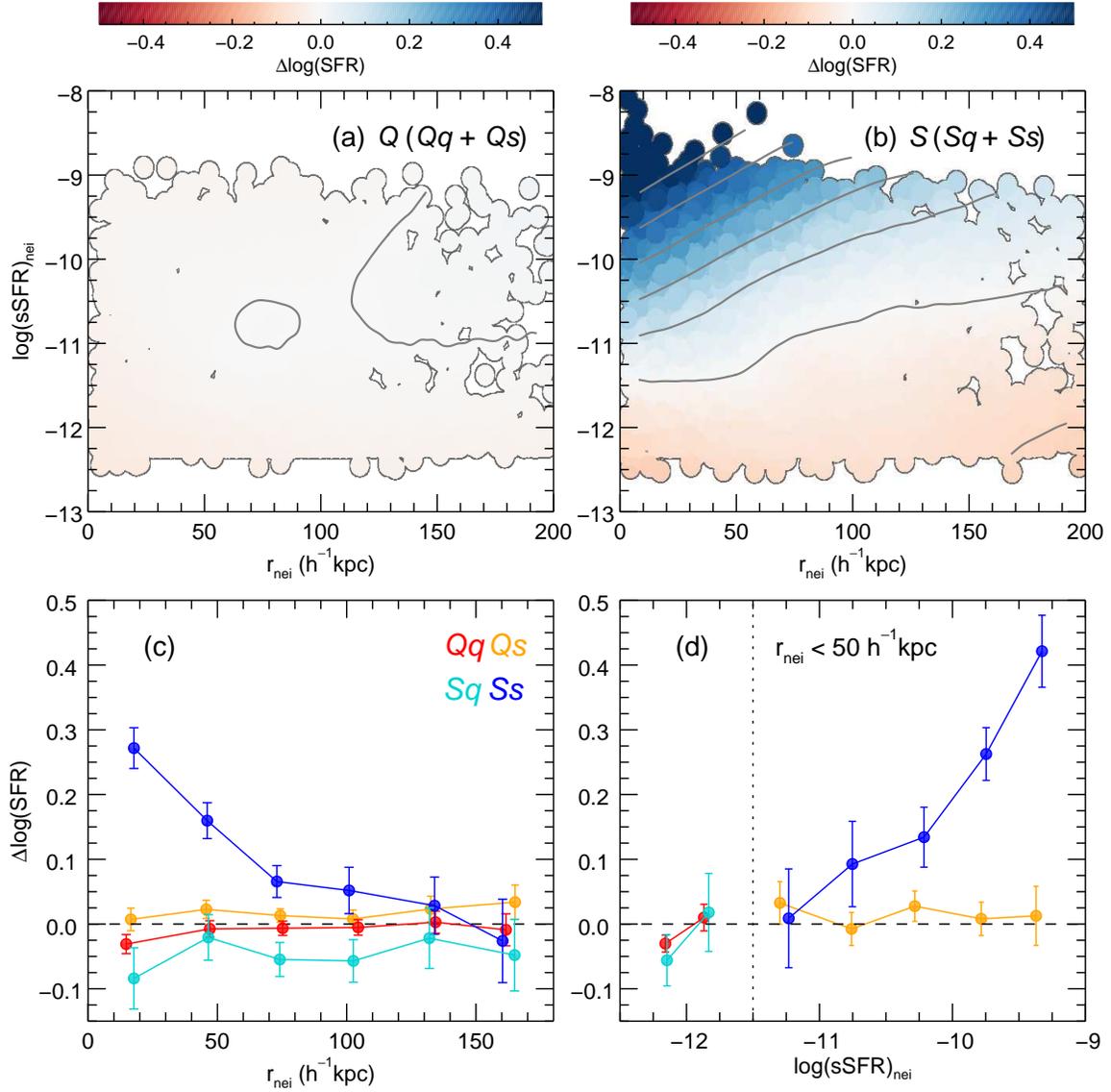}
\caption{Interaction-induced changes in the SFR as functions of the distance to the nearest neighbor and the sSFR of the nearest neighbor. Top panels: 2D trends obtained by the local regression algorithm of \citet{2013MNRAS.432.1862C} for (a) quiescent and (b) star-forming paired galaxies, respectively. The stronger the SF is induced by interactions, the bluer the color is. Contours are drawn at the interval of 0.1 dex in $\Delta \textrm{log}(\textrm{SFR})$. Bottom panels: mean interaction-induced SF as functions of (c) the projected distance to the nearest neighbor and (d) the sSFR of the nearest neighbor. Panel (d) restricts the sample to close paired galaxies with a separation of less than 50 $h^{-1}$kpc. Four possible combinations of SF activities for galaxy pairs are shown in different colors: \grp{Qq} (red), \grp{Qs} (orange), \grp{Sq} (cyan), \grp{Ss} (blue). In order to construct the control samples, for each target galaxy in the pair sample 1000 isolated galaxies in the same SF category as the target galaxy are randomly drawn. The control sample for star-forming (quiescent) pairs only consists of star-forming (quiescent) isolated galaxies. Error bars represent the standard error of the mean. \label{fig:ssfrsubs}}
\end{figure*}

The quenching induced by galaxy interactions was reported in previous studies based on the SDSS. For example, \citet{2009MNRAS.399.1157P} found an excess of red galaxies in the pair sample, particularly at high-density environments. \citet{2015MNRAS.450.1546B} found that galaxies with stellar masses between 10$^{10}$ and 10$^{10.5}$ $M_{\mathrm{sun}}$ interacting with a close pair show a lower star-forming fraction than the average. Our finding suggests that the reduction of the star-forming fraction in paired galaxies is in fact led by the \grp{q} group and that the star-forming fraction of paired galaxies is linked to the SF status of interacting neighbors.

\subsection{Interaction-induced ``Enhancement'' in the Star Formation Rate of the \grp{s} Group} \label{subsec:ssfr}

Now we examine the interaction-induced SFR in the \grp{q} and \grp{s} groups. The interaction-induced SFR is defined as the difference in the SFR between the pair sample and the control sample such that
\begin{equation}
\Delta \textrm{log}(\textrm{SFR}) = \textrm{log}(\textrm{SFR}_{\mathrm{pair}}) - \textrm{log}(\textrm{SFR}_{\mathrm{control}}). 
\end{equation}
We use the SFR measured within the fiber instead of the aperture-corrected global SFR because the control sample has the same redshift distribution as the pair sample, so that a fair comparison is guaranteed between the fiber SFRs of the pair and control samples. The reason to use the SFR instead of the sSFR is to avoid uncertainty in estimating mass within the fiber. We tested sSFR within the fiber, but it did not change the conclusion. In this section, we take into account the SF activities of target galaxies themselves, as well as those of their neighbors. This leads to four distinct groups; \grp{Qq}, \grp{Qs}, \grp{Sq}, and \grp{Ss}. For a fair comparison, the control samples for the \grp{Qq} and \grp{Qs} (\grp{Sq} and \grp{Ss}) groups are constructed only using quiescent (star-forming) isolated galaxies. Thus, $\Delta \textrm{log}(\textrm{SFR})$ for the \grp{Qq} and \grp{Qs} (\grp{Sq} and \grp{Ss}) groups means the change of the SFR with respect to the isolated quiescent (star-forming) galaxies, and thus it is not affected by the reduction of the SF fraction discussed in Section \ref{subsec:sffrac}. The construction of the control sample is repeated 1000 times like before, so each galaxy in the pair sample has 1000 corresponding control galaxies, and then the mean values and standard errors of $\Delta \textrm{log}(\textrm{SFR})$ are calculated at each bin.

\begin{figure*}[t]
\epsscale{0.97}
\plotone{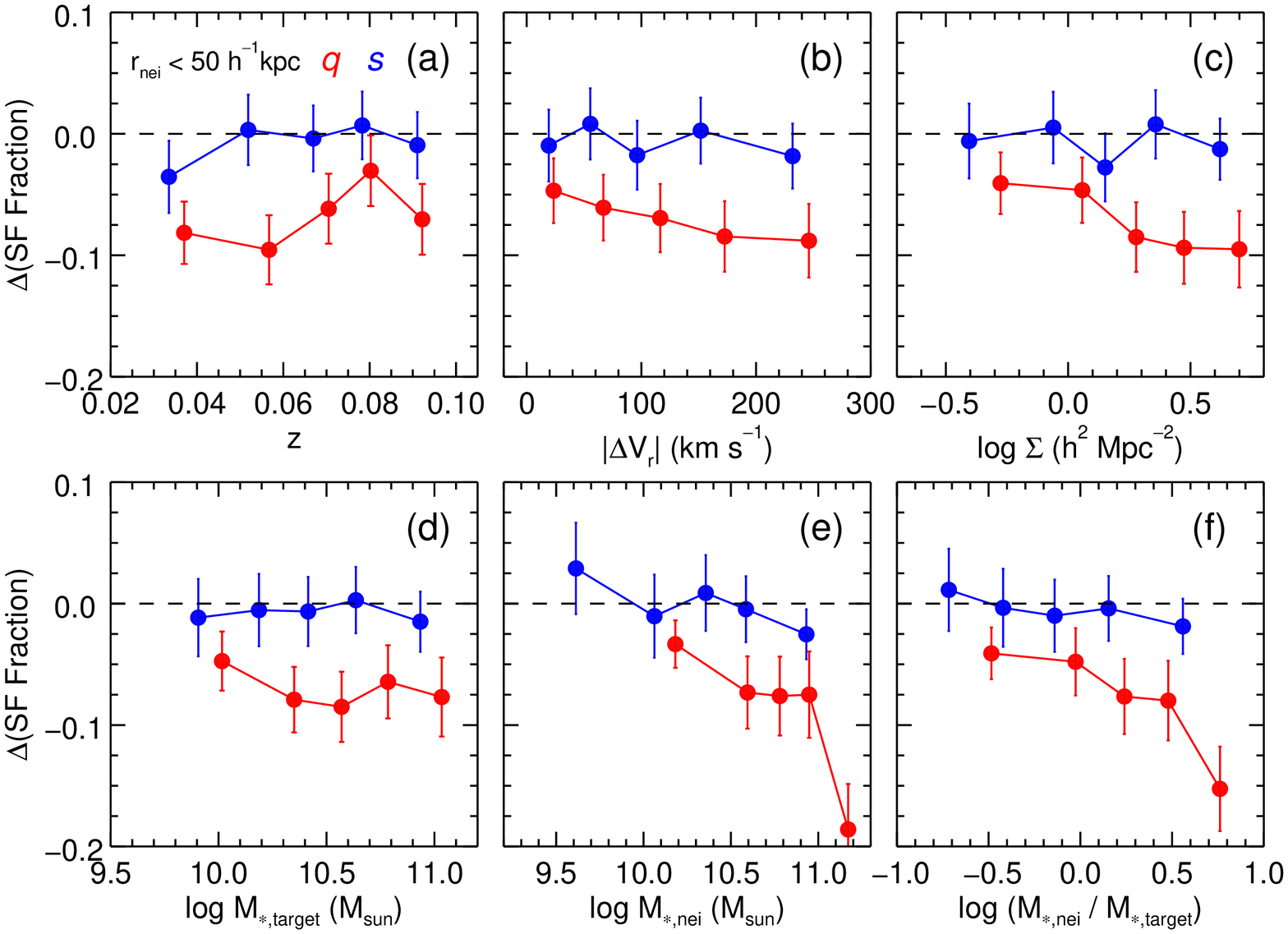}
\caption{Interaction-induced changes in the number fraction of star-forming galaxies with respect to (a) the redshift, (b) the relative radial velocity between the target and neighbor, (c) the local density parameter, (d) the stellar mass of the target, (e) the stellar mass of the neighbor, and (f) the mass ratio between the neighbor and target. The sample is restricted to close paired galaxies with a projected separation less than 50 $h^{-1}$kpc. The fractions of the star-forming galaxies in the \grp{q} and \grp{s} groups are shown as red and blue circles, respectively. For each group, five bins contain an equal number of galaxies. Error bars represent the standard error. \label{fig:depend_frac}}
\end{figure*}

Figure \ref{fig:ssfrsubs} presents the interaction-induced SF as a function of the projected separation from the nearest neighbor and the sSFR of the neighbor. Panels (a) and (b) show in the 2D space $\Delta \textrm{log}(\textrm{SFR})$ for the \grp{Q} group (i.e., quiescent galaxies paired with either quiescent or star-forming neighbors) and the \grp{S} group (i.e., star-forming galaxies paired with either quiescent or star-forming neighbors), respectively. In order to reduce the stochastic noise and to bring out the general trend, we apply the \texttt{CAP\_LOESS\_2D} routine of \citet{2013MNRAS.432.1862C}, which implements the multivariate LOESS (locally weighted regression) algorithm of \citet{1988JASA..83...596C}. In panel (a), the interaction-induced SF for the \grp{Q} group is only marginal regardless of the separation and the sSFR of the neighbor. On the other hand, for the \grp{S} galaxies in panel (b), the SFR increases over isolated star-forming galaxies when the separation decreases and the sSFR of the neighbor increases. Particularly at small separations, the sSFR of the neighbor is even more important than the separation.

Figure \ref{fig:ssfrsubs}(c) shows $\Delta \textrm{log}(\textrm{SFR})$ as a function of the projected separation for the four distinct groups. Since we restrict quiescent galaxies to galaxies with log(sSFR) $<$ $-$11.5 by definition, the mean SFR of the \grp{Qq} (red line) and \grp{Qs} (orange) groups barely changes regardless of the separation. The mean SFR of \grp{Sq} galaxies (cyan), even when they have a close quiescent neighbor, is not much different from the control sample. The \grp{Sq} galaxies, in fact, show a little depression in the SFR. By contrast, the mean SFR of \grp{Ss} galaxies (blue) increases with the decreasing separation, showing twice higher SFR at $r_{\mathrm{nei}}$ $<$ 30 $h^{-1}$kpc than the control sample. Our findings agree with \citet{2016ApJS..222...16C}, who examined the difference in the interaction-induced SF between the quiescent and star-forming galaxies based on far-infrared observations of close major-merger pairs. We confirm, using a larger sample of galaxies with a strict control sample, that the interaction-induced SF is determined by the gas property of the interacting companion. If the tidal effect, as often interpreted, drives the interaction-induced SF, there is no reason to have a different value of $\Delta \textrm{log}(\textrm{SFR})$ between the \grp{q} and \grp{s} groups. Moreover, neighbors are on average more massive in the \grp{q} group than in the \grp{s} group, so the strength of tidal force exerted by neighbors is usually larger in the \grp{q} group.

Figure \ref{fig:ssfrsubs}(d) shows the interaction-induced SF for close pairs as a function of the sSFR of the neighbor. We restrict the sample to galaxies with a separation of less than 50 $h^{-1}$kpc. Since we restrict quiescent galaxies to galaxies with log(sSFR) $<$ $-$11.5 by definition, the mean SFR of the \grp{Qq} (red line) and \grp{Qs} (orange) groups barely changes with the sSFR of the neighbor. The SFR of the \grp{Sq} group (cyan) is not enhanced even though they have a close neighbor, and slight reduction of SFR is seen when having fully quenched neighbors (i.e., log(sSFR) $<$ $-$12). Remarkably, for the \grp{Ss} group (blue), the strength of SF rises with the sSFR of the neighbor. The SFR of star-forming galaxies having a neighbor with log(sSFR) $>$ $-$9.5 increases up to about 3 times that of the control sample. We should note that the massive SF found in the \grp{Ss} group may be the consequence of mutual interaction between two galaxies. If two galaxies in a pair may exert strong tidal influence on each other, the interaction can induce high SFRs in both galaxies. Nevertheless, it is difficult to explain with the tidal impact alone the fact that there is no enhancement of SF in the \grp{Sq} group (cyan), where the target galaxies themselves contain enough gas to fuel new SF. Thus, our results suggest that, contrary to the usual picture that the SF enhancement is governed by the tidal effect, hydrodynamic mechanisms also play an essential role during galaxy interactions.

\section{Discussion} \label{sec:discuss}

\subsection{Is the Difference between the \grp{q} and \grp{s} Groups Real?} \label{subsec:depend}

\begin{figure*}[t]
\epsscale{0.97}
\plotone{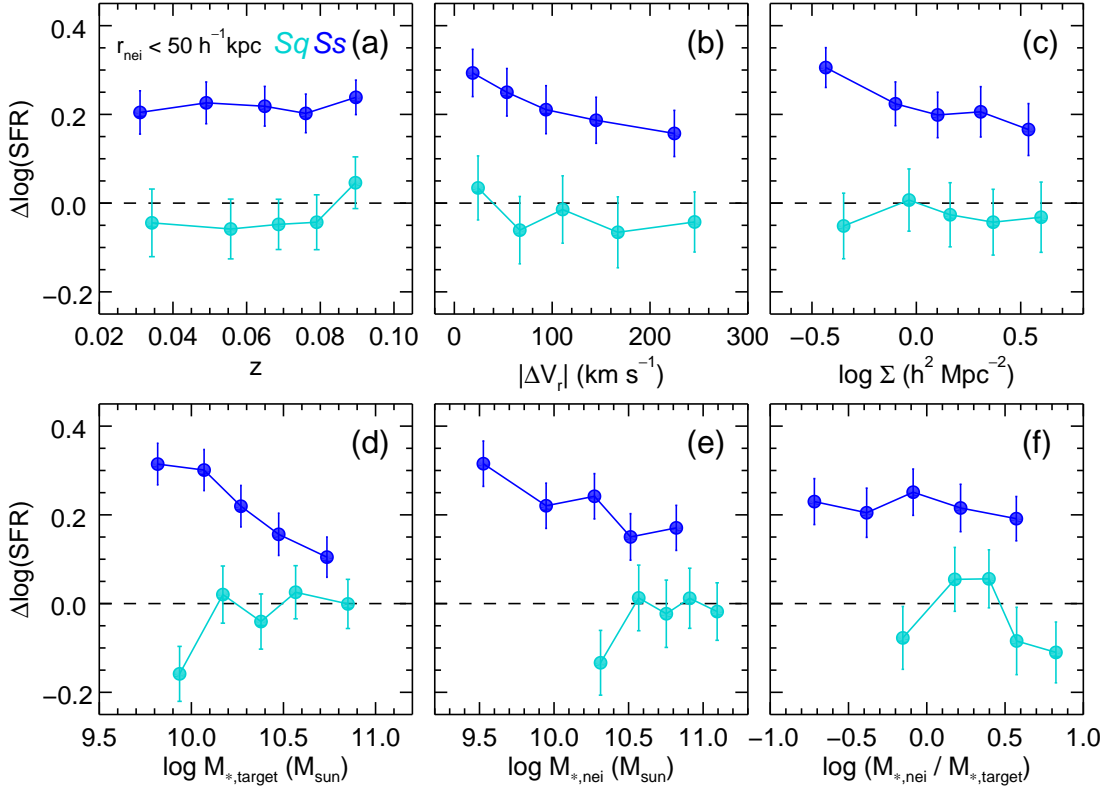}
\caption{Same as Figure \ref{fig:depend_frac}, but for interaction-induced changes in the SFR. \label{fig:depend_sfr}}
\end{figure*}

We have shown that the interaction-induced variation in the fraction of star-forming galaxies and the SFR are strong functions of the separation to the neighbor and the neighbor's SFR. The strength of the interaction-induced SF is also known to depend on various parameters, such as redshift, target galaxies' mass, and local environment. For instance, it is known that the interaction-induced SF is usually prominent only in low-density environments \citep{2010MNRAS.407.1514E}. Since the \grp{q} galaxies tend to reside in denser environments than the \grp{s} galaxies on average, the difference in $\Delta \textrm{log}(\textrm{SFR})$ between the \grp{q} and \grp{s} groups may be caused by their different environments, not by the neighbor's SF property.

In Figure \ref{fig:depend_frac}, we examine the effect of various parameters on the interaction-induced reduction of star-forming galaxy fraction. We use only close paired galaxies having a companion within a projected separation of 50 $h^{-1}$kpc to see a clear trend. In each panel, the red and blue circles represent the behaviors of the \grp{q} and \grp{s} galaxies, respectively. In all panels, the star-forming galaxy fraction in the \grp{q} group with respect to the control sample is always negative, regardless of the parameters, while the fraction in the \grp{s} group remains similar to the control sample. Panel (a) shows the dependence on the redshift. The reduction of star-forming galaxies only appears in the \grp{q} group, with no clear trend on the redshift. Panel (b) shows the dependence on the radial velocity difference between the target and the neighbor. In the \grp{q} group, the number of star-forming galaxies slightly decreases as the relative velocity increases, but the \grp{s} group always shows a similar fraction to the control sample. Panel (c) shows the dependence on the local density. The \grp{q} galaxies in a denser environment are less star-forming, but the \grp{s} group does not show such a trend. Panel (d) shows the dependence on the stellar mass of the target galaxy. The reduction of star-forming galaxies only appears in the \grp{q} group, with no clear trend on target galaxies' mass. Panels (e) and (f) show the dependence on the mass of the neighbor and the mass ratio between the target and the neighbor, respectively. For the \grp{q} group, the reduction in the number of star-forming galaxies is more severe when a galaxy has a more massive neighbor.

In summary, there is a significant reduction of star-forming galaxy fraction in the \grp{q} group regardless of the redshift, relative velocity, local environment, stellar mass, and mass ratio, while the fraction in the \grp{s} group remains similar to the control sample. The more effective quenching by more massive neighbors in the \grp{q} group can be explained if the quenching results from tidal stripping and heating by the neighbor \citep{2015MNRAS.452..616D} or from the cutoff of gas accretion by a hot gas halo of the neighbor \citep{2015MNRAS.447..374G}.

\begin{figure*}[t]
\epsscale{1.0}
\plotone{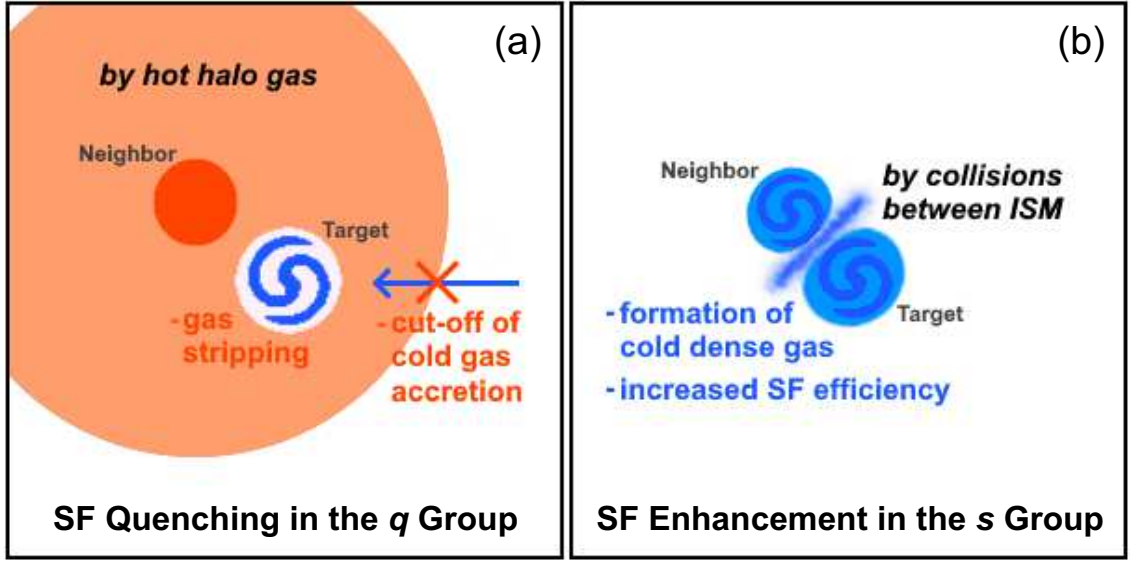}
\caption{Schematic diagrams illustrating mechanisms that explain the observational results (a) for the \grp{q} group and (b) for the \grp{s} group. \label{fig:mechanisms}}
\end{figure*}

Figure \ref{fig:depend_sfr} shows the effect of various parameters on the interaction-induced SF. Figure \ref{fig:depend_sfr} shows the effect of various parameters on the interaction-induced SF. In each panel, the cyan and blue circles represent galaxies in the \grp{Sq} and \grp{Ss} groups, respectively. In all panels, the SFR in the \grp{Ss} group with respect to the control sample is always positive, regardless of the parameters associated with the interaction. Unlike the \grp{Ss} group, the SF enhancement is not detected for the \grp{Sq} group regardless of the parameters. Panel (a) shows that the SF enhancement only appears in the \grp{Ss} group, with no clear dependence on the redshift. Panel (b) shows that $\Delta \textrm{log}(\textrm{SFR})$ of the \grp{Ss} group decreases as the relative velocity increases, most likely because a large relative velocity raises the possibility of contamination by false projected pairs. The \grp{Sq} galaxies do not exhibit the SF enhancement for all relative velocities. Panel (c) shows that $\Delta \textrm{log}(\textrm{SFR})$ of the \grp{Ss} sample shrinks at denser environments, but the contrast between the \grp{Sq} and \grp{Ss} groups is significant at all environments. In panel (d), the SF enhancement is more significant for lower-mass \grp{Ss} galaxies, and the SF activity of the \grp{Sq} galaxies is depressed at the lowest stellar mass bin. In panel (e), the trend is basically similar to panel (d) because the mass of the neighbor is restricted to be 0.1\,--\,10 times that of the target. Panel (f) shows that, although galaxies in the \grp{Sq} group generally interact with more massive neighbors than those in the \grp{Ss} group, the SF enhancement only appears in the \grp{Ss} group for fixed mass ratio. 

In summary, the SFR of the \grp{Sq} group is similar to or slightly lower than that of the control sample, regardless of the redshift, relative velocity, local environment, stellar mass, and mass ratio. The SFR in the \grp{Ss} group with respect to the control sample is always positive, regardless of the parameters associated with the interaction. It appears that the most important parameter that distinguishes the \grp{Sq} and \grp{Ss} groups is the SFR of the neighbor.

\subsection{What Mechanisms Are behind the Difference between the \grp{q} and \grp{s} Groups?} \label{subsec:mechanisms}

This section addresses what mechanisms are involved in the difference between the \grp{q} and \grp{s} groups. We have shown (a) that quiescent neighbors quench the SF, and even more so when neighbors are closer and more quiescent, and (b) that star-forming neighbors efficiently enhance SF, and even more so when neighbors are closer and more star-forming. The results point to a pivotal role of hydrodynamic mechanisms during galaxy interactions. Figure \ref{fig:mechanisms} schematically shows the possible scenarios explaining the observational results.

Figure \ref{fig:mechanisms}(a) illustrates a hydrodynamic mechanism that ensures SF quenching of galaxies paired with quiescent neighbors. The influence of the hot gaseous medium surrounding quiescent neighbors can help to quench SF. The fraction of star-forming galaxies for the \grp{q} group is lower than that for the control sample (Figure \ref{fig:sffrac}), and the reduction extends out to 150 $h^{-1}$kpc, which is the typical size of the virial radius for Milky Way--sized galaxies. The fact that the \grp{q} galaxies with more massive neighbors are more quenched supports the notion of the hot halo impact (Figure \ref{fig:depend_frac}) because a galaxy should exceed a critical mass to maintain a stable hot gas halo \citep{2018MNRAS.473..538C}. The hot gas halo is able to shut off the additional cold gas supply and strip gas material from galaxies traveling through the hot halo \citep[see also][]{2009ApJ...691.1828P, 2011A&A...535A..60H}. When a galaxy infalls into the hot halo, the galaxy slowly consumes its gas content over a few gigayears, and then the SF gets rapidly quenched in a short period of time \citep{2013MNRAS.432..336W, 2019MNRAS.486.5184R}. This explains the SFR distribution of the \grp{q} group that has a higher fraction of quiescent galaxies with little change in the mean SFR of star-forming galaxies (Figure \ref{fig:ssfrhist}).

Contrary to the usual expectation that even quiescent neighbors enhance SF via the tidal effect, the induced SF is not detected in interactions with quiescent neighbors. It seems that the recent enhancement of the SFR in the \grp{Sq} galaxies by the tidal effect is canceled out by the gas depletion owing to the hot gas halo of the neighbor (Figure \ref{fig:ssfrsubs}). In theoretical models of galaxy mergers, the hot halo component of galaxies was often omitted. Only a few authors have tested the effect of the hot halo on interaction-induced SF \citep{2011MNRAS.415.3750M, 2012MNRAS.423.2045M, 2015ApJ...805..131H}, and they concluded in common that the hot halo affects the interaction-induced SF. However, examining hot halos in theoretical models is still hard, since the interplay between baryonic physics (e.g., cooling process) and hydrodynamic processes (e.g., shocks) is too subtle.

Figure \ref{fig:mechanisms}(b) illustrates a hydrodynamic mechanism that ensures SF enhancement of galaxies paired with star-forming neighbors. The interaction between the ISMs of two galaxies in a pair can take a role in enhancing SF. It has recently been under debate whether the interaction-induced SF is achieved by an enhanced efficiency of SF or an increase of the gas mass \citep[see, e.g.,][]{2018ApJ...868..132P}. Whatever the exact origin is, the process seems more complex than the conventional view that centrally concentrated material ignites the SF. Regarding an enhanced efficiency of SF, on the one hand, \citet{2018ApJS..237....2Z} reported that paired galaxies interacting with a spiral neighbor show a higher SFR per unit H \textsc{\romannumeral 1} gas mass (i.e., a higher SF efficiency) than galaxies interacting with an elliptical neighbor. The difference in the SF efficiency suggests that hydrodynamic interactions, i.e, the ISM collision between two galaxies in an \grp{Ss} pair, play a role in enhancing SF \citep[see also][]{2016ApJS..222...16C, 2018ApJS..237....2Z}. Regarding an increase of the gas mass, on the other hand, \citet{2018ApJ...868..132P} found that there is an enhancement of molecular gas mass in interacting galaxies, while the SFR per unit molecular gas mass is unchanged. Recent merger simulations conducted by \citet{2019MNRAS.485.1320M} showed that cold dense gas is supplied by the cooling process during the early phase of interactions. While the tidal torque is too weak to produce the radial gas infall in this `galaxy pair' stage \citep{2015MNRAS.446.2038R}, more efficient is the cold gas supply by hydrodynamic compression through the ISM collision between two gas-rich galaxies in \grp{Ss} pairs. In this regard, why the SF enhancement is not detected in the \grp{Sq} galaxies is explained by the absence of the ISM to collide, as well as the hot gas halo effect (Figure \ref{fig:ssfrsubs}).

We expect that cosmological simulations can provide valuable constraints on the possible mechanism behind the phenomenon. A comprehensive picture of galaxy mergers including cold gas accretion, minor mergers with third bodies, and non-ad-hoc orbital parameters can be accomplished only in the cosmological context. Obviously, the \grp{Sq} and \grp{Ss} pairs occupy different contexts in the cosmic web. For instance, \citet{2018A&A...619A..24M} showed, using the SDSS, that S+E pairs are more aligned to nearby filaments and reside in more massive filaments than S+S pairs. \citet{2018MNRAS.480.2266M} found that disk galaxies at $z = 0$ have had more prograde mergers than spheroids over cosmic time in a cosmological hydrodynamical simulation. Some recent studies have started to investigate the SF enhancement in paired galaxies within cosmological simulations. \citet{2015MNRAS.452.2845K} showed that the SF enhancement during major mergers is also detectable in a cosmological simulation, albeit weaker than observed. \citet{2018MNRAS.479.3381B} investigated galaxy mergers based on cosmological zoom-in simulations and showed that the SF enhancement is comparable to observations and idealized merger simulations. While the resolution of state-of-the-art cosmological hydrodynamic simulations (the spatial resolution of 0.1--1 kpc) is still too low compared to that of flagship merger simulations (1--10 pc), the zoom-in technique has been a good compromise to explore the interaction-induced variation in the cosmological context \citep[e.g.,][]{2016MNRAS.462.2418S, 2018MNRAS.479.3381B, 2018MNRAS.475.1160H}. A detailed analysis of paired galaxies in cosmological simulations, including the role of neighbors in enhancing SF, is yet to be conducted in future researches.

\section{Conclusion} \label{sec:conclusion}

We have investigated the effect of the nearest neighbor on the SF activity based on a large sample of paired galaxies. Galaxies paired with neighbors are selected from the SDSS and classified according to the SF activities of its own and of interacting neighbors. To minimize the selection bias, we carefully construct the control sample of isolated counterparts, against which we detect purely interaction-induced changes in paired galaxies. In particular, the paired galaxies have been selected by strict criteria: $r_{\mathrm{nei}}$ $<$ 200 $h^{-1}$kpc, $|\Delta$V$_{\mathrm{r}}|$ $<$ 300 km\,s$^{-1}$, 0.1 $<$ ($M_{\mathrm{*,nei}}$/$M_{\mathrm{*,target}}$) $<$ 10, and a neighbor should be the nearest and the most influential (highest $\Theta_{\mathrm{nei}}$) one (Section \ref{sec:sample}). Our sample consists of 5024 galaxies interacting with a quiescent neighbor (the \grp{q} group) and 5118 galaxies interacting with a star-forming neighbor (the \grp{s} group). The control sample is built based on the redshift, stellar mass, and local density (Section \ref{sec:control}). Based on the SDSS and MPA-JHU catalogs, we derive the enhanced or reduced SF of galaxies in pairs with respect to the control sample (Section \ref{sec:result}). Our results are summarized as follows:

\begin{enumerate}

\item{The sSFR distribution of the pair sample is more dispersed than the isolated control sample. Once the pair sample is divided into the \grp{q} (i.e., galaxies interacting with a quiescent neighbor) and \grp{s} (i.e., galaxies interacting with a star-forming neighbor) groups, the increase in the number of quiescent galaxies is led by the \grp{q} group and the enhancement of the SFR is led by the \grp{s} group (Section \ref{sec:result}).}

\item{The number fraction of star-forming galaxies decreases in the \grp{q} group compared to the control sample, and this is more so when quiescent neighbors are closer and more quiescent. The decrease extends out to the separation of 150 $h^{-1}$kpc. In contrast, the star-forming fraction of the \grp{s} group does not show a deviation from that of the control sample (Section \ref{subsec:sffrac}).}

\item{The \grp{Sq} group (i.e., star-forming galaxies interacting with a quiescent companion) does not show the SF enhancement induced by interactions. Only for the \grp{Ss} group (i.e., star-forming galaxies interacting with a star-forming companion) is the mean SFR significantly enhanced compared to that of isolated star-forming galaxies. The interaction-induced SF of the \grp{Ss} galaxies increases as the separation decreases and the sSFR of the neighbor increases (Section \ref{subsec:ssfr}).}

\item{The differences in the star-forming galaxy fraction and the SFR between the \grp{s} group and the \grp{q} group do not stem from different galaxy properties such as the redshift, relative velocity, local environment, stellar mass, and mass ratio. The most important parameter that distinguishes the \grp{q} and \grp{s} groups is the SFR of the neighbor (Section \ref{subsec:depend}).}

\item{Our findings, especially the intimate connection of SF to the status and strength of neighbors' SF, suggest the crucial role of the hydrodynamic effect on the induced SF during galaxy--galaxy interactions (Section \ref{subsec:mechanisms}). It is plausible that the underlying mechanisms are (a) the existence of the hot gas halo in the \grp{q} group and (b) the collision between two gas-rich disks in the \grp{s} group.}


\end{enumerate}

\acknowledgments
{
This work is supported by the Mid-career Researcher Program (No. 2019R1A2C3006242) and the SRC Program (the Center for Galaxy Evolution Research; No. 2017R1A5A1070354) through the National Research Foundation of Korea. 

Funding for the SDSS and SDSS-II has been provided by the Alfred P. Sloan Foundation, the Participating Institutions, the National Science Foundation, the U.S. Department of Energy, the National Aeronautics and Space Administration, the Japanese Monbukagakusho, the Max Planck Society, and the Higher Education Funding Council for England. The SDSS website is http://www.sdss.org/.

The SDSS is managed by the Astrophysical Research Consortium for the Participating Institutions. The Participating Institutions are the American Museum of Natural History, Astrophysical Institute Potsdam, University of Basel, University of Cambridge, Case Western Reserve University, University of Chicago, Drexel University, Fermilab, the Institute for Advanced Study, the Japan Participation Group, Johns Hopkins University, the Joint Institute for Nuclear Astrophysics, the Kavli Institute for Particle Astrophysics and Cosmology, the Korean Scientist Group, the Chinese Academy of Sciences (LAMOST), Los Alamos National Laboratory, the Max-Planck-Institute for Astronomy (MPIA), the Max-Planck-Institute for Astrophysics (MPA), New Mexico State University, Ohio State University, University of Pittsburgh, University of Portsmouth, Princeton University, the United States Naval Observatory, and the University of Washington.
}

\end{document}